\newcommand{\nhu}{\,cm$^{-2}$\,}

\newcommand{\lu}{\,erg s$^{-1}$\,}
\newcommand{\fu}{\,erg cm$^{-2}$ s$^{-1}$\,}
\newcommand{\lx}{$L_{\rm x}$}
\newcommand{\xmm}{{\em XMM-Newton}}
\newcommand{\hd}{HD\,17156}

\documentclass[iop,superscriptaddress]{emulateapj}
\usepackage{natbib}
\usepackage{epstopdf}
\received{2015 June 23}
\accepted{2015 August 18}

\shortauthors{Maggio \& et al.}
\shorttitle{Star-Planet Interaction in HD 17156}

\begin{document}

\title{Coordinated X-ray and Optical observations of Star-Planet
Interaction in HD 17156}

%

\author{A. Maggio}
\affiliation{INAF -- Osservatorio Astronomico di Palermo,\\
piazza del Parlamento 1, I--90134 Palermo, Italy;\\
{\em maggio@astropa.unipa.it}}
\author{I. Pillitteri}
\affiliation{SAO--Harvard Center for Astrophysics, Cambridge, MA 02138, USA}
\affiliation{INAF -- Osservatorio Astronomico di Palermo, Italy}
\author{G. Scandariato, A.F. Lanza} 
\affiliation{INAF -- Osservatorio Astrofisico di Catania, Italy}
\author{S. Sciortino} 
\affiliation{INAF -- Osservatorio Astronomico di Palermo, Italy}
\author{F. Borsa} 
\affiliation{INAF -- Osservatorio Astronomico di Brera, Milano, Italy}
\author{A.S. Bonomo}
\affiliation{INAF - Osservatorio Astrofisico di Torino, Pino Torinese, Italy}
\author{R. Claudi}
\affiliation{INAF -- Osservatorio Astronomico di Padova, Italy}
\author{E. Covino}
\affiliation{INAF -- Osservatorio Astronomico di Capodimonte, Italy}
\author{S. Desidera, R. Gratton}
\affiliation{INAF -- Osservatorio Astronomico di Padova, Italy}
\author{G. Micela}
\affiliation{INAF -- Osservatorio Astronomico di Palermo, Italy}
\author{I. Pagano}
\affiliation{INAF -- Osservatorio Astrofisico di Catania, Italy}
\author{G. Piotto}
\affiliation{INAF -- Osservatorio Astronomico di Padova, Italy}
\affiliation{Dip. di Fisica e Astronomia G. Galilei -- Universit\`a di Padova, Italy}
\author{A. Sozzetti}
\affiliation{INAF - Osservatorio Astrofisico di Torino, Pino Torinese, Italy}
\author{R. Cosentino}
\affiliation{Fundaci\'on Galileo Galilei -- INAF, Bre\~{n}a Baja, TF--Spain}
\author{J. Maldonado}
\affiliation{INAF -- Osservatorio Astronomico di Palermo, Italy\\
\\}

\begin{abstract}

The large number of close-in Jupiter-size exoplanets 
prompts the question whether star--planet interaction (SPI) effects
can be detected. We focused our attention on the system \hd,
having a Jupiter-mass planet in a very eccentric
orbit.
Here we present results of the \xmm\ observations and of a 
five months coordinated optical campaign with the HARPS-N 
spectrograph\footnote{Based on observations collected at the Italian Telescopio
Nazionale Galileo (TNG), operated on the island of La Palma by
the Fundanci\'on Galileo Galilei of the INAF (Istituto Nazionale di
Astrofisica), in the frame of the programme {\it Global Architecture of
Planetary Systems} (GAPS).}.
We observed \hd\ with \xmm\
when the planet was approaching the apoastron and then
at the following periastron passage, quasi simultaneously with HARPS-N. 
We obtained a clear ($\approx 5.5\sigma$) X-ray detection
only at the periastron visit, accompanied by a significant increase of
the $R'_{\rm HK}$ chromospheric index.
We discuss two possible scenarios for the activity enhancement: 
magnetic reconnection and flaring or accretion onto the star 
of material tidally stripped from the planet. In any case,
this is possibly the first evidence of a magnetic SPI effect caught in action.

\end{abstract}


\clearpage

\section{Introduction}
\label{sec:intro}

The discovery of new extrasolar planetary systems is
today routinely carried out with a wealth of space-based and ground-based 
observational facilities, and our knowledge frontier has already moved on 
from their mere search to
the determination of stellar and planetary properties which are relevant 
to understand the evolution of these bodies during stellar lifetime. 
In particular, one of the open issues is the extent of 
magnetic star--planet interaction
(SPI) and its detectable effects in systems with close-in massive planets 
(so called ``hot Jupiters''). This phenomenon is astrophysically
important for the characterization of planetary magnetospheres, and as a
mechanism of energy and angular momentum transfer between the host star
and its planet(s) \citep{Cuntz2000}.

Observations of variable chromospheric emission signatures,
such as the Ca II H\&K lines cores, were reported for
a few stars, with a modulation period close to the orbital period of the planet
\citep{Shkolnik2005,Shkolnik2008}, 
and interpreted as evidence of SPI of magnetic origin. 
In fact, a tidally induced effect can be excluded because -- for
low-eccentricity orbits -- the periodic
modulation is expected twice per orbital period, due to the two 
tidal bulges on
the opposite sides of the stellar surface. 
Conversely, a purely stellar activity effect is unlikely
because  the relevant time scale should be the stellar rotation period rather than
the planetary orbital period. 

However, significant chromospheric variability is visible only at some
epochs, and the existence of detectable SPI signatures at X-ray
wavelengths is a matter of debate. \citet{Kashyap2008}
showed that stars with hot Jupiters are statistically brighter in X-rays
than stars without hot Jupiters, and \citet{Scharf2010} claimed a correlation
between the X-ray luminosity of the host star and the planetary mass
which he ascribed to the same phenomenon. Detailed statistical studies
by \citet{Poppenhaeger2010} and \citet{Poppenhaeger2011},
reached instead the conclusion that the above results can be
affected by observational biases, and that SPI induces quite small
effects in most cases, which become important and measurable only in
peculiar systems. 

Recently, \citet{Pillitteri2010} and \citet{PoppenhaegerWolk2014}
pinpointed the case of two wide binaries where the planet-hosting primary
component is much brighter in X-rays than expected by considering the
activity level of the coeval companion. 
This finding suggests that SPI, by acting through transfer of angular momentum
from the planet to the star and increasing thus the stellar rotation and activity, 
may lead to enhanced X-ray emission and thus ``rejuvenating'' the star.

Detailed studies of individual systems were also performed, leading
to different results: the occurrence of repeated flaring events in
HD\,189733\,A just after egress from planet occultation, observed
with \xmm\ \citep{Pillitteri2011,Pillitteri2014,Pillitteri2015},
suggests a systematic SPI that leads to X-ray and FUV
variability phased with the planetary motion; instead, a long \xmm\
monitoring program of HD\,179949 \citep{Scandariato2013},
showed evidence of chromospheric and coronal variability most likely
due to stellar rotation and to intrinsic short-term activity evolution,
but no clear signature related to the orbital motion of the planet.

\begin{deluxetable}{lrcrr}
\tablewidth{0pt}
\tablecaption{Observation Log
\label{tab:obslog}}
\tablecolumns{9}
\tablehead{
\multicolumn{1}{c}{Date} & & 
\multicolumn{1}{c}{Orbital} & \multicolumn{1}{c}{$t_{\rm exp}$} & \\
\multicolumn{1}{c}{(2014)} & \multicolumn{1}{c}{$MJD^a$} &
\multicolumn{1}{c}{Phase} & \multicolumn{1}{c}{[s]}
}
\startdata
\multicolumn{4}{c}{\xmm Observations} & $\log L_{\rm x}/L_{\rm bol}$ \\
Sep  5 & 56905.02595 & 0.267--0.285 & 32,100 & $< -7.3$ \\
Sep 20$^b$ & 56920.96779 & 0.018--0.037 & 32,900 & $-7.0$ \\
\hline
\multicolumn{4}{c}{HARPS-N Observations} & 
\multicolumn{1}{c}{$\log R'_{\rm HK}$}\\
Aug  5 & 56874.23161 & 0.815 & 900 & $-5.079 \pm 0.006$ \\
Aug  7 & 56876.23117 & 0.909 & 900 & $-5.083 \pm 0.007$ \\
Aug  8 & 56877.21796 & 0.956 & 900 & $-5.097 \pm 0.005$ \\
Aug  9 & 56878.23348 & 0.004 & 900 & $-5.096 \pm 0.003$ \\
Aug 21$^c$ & 56890.24703 & 0.570 & 900 & $-5.080 \pm 0.004$ \\
Sep  9 & 56909.07199 & 0.457 & 10,800 & $-5.094 \pm 0.001$ \\
Sep 21$^b$ & 56921.20339 & 0.029 & 7,200 & $-5.077 \pm 0.002$ \\
Nov 10 & 56971.08149 & 0.380 & 7,200 & $-5.096 \pm 0.002$ \\
Nov 25 & 56986.07491 & 0.087 & 900 & $-5.098 \pm 0.004$ \\
Dec  7 & 56998.03466 & 0.650 & 900 & $-5.095 \pm 0.008$ \\
Dec 14 & 57005.08871 & 0.983 & 7,200 & $-5.100 \pm 0.004$ \\
\enddata
\tablecomments{$a$: $=HJD-2400000.5$. $b$: simultaneous observations.
$^c$ High umidity night ($\sim 80$\%).
}
\end{deluxetable}

From a theoretical point of view, analytical models were developed to
explain the periodic chromospheric enhancements as due to flare-like
magnetic reconnection events \citep[e.g.,][]{Lanza2009}, with
an expected dependence on the relative velocity between the coronal and
planetary magnetic fields.  Another key parameter is given by the relative
strength of planetary and stellar magnetic fields, which determines
the level of intersection of the respective Alfv\'enic surfaces, as
investigated by \citet{Cohen2009} and \citet{Lanza2012}.
Sophisticated MHD simulations by \citet{Cohen2009}
confirm that the interaction of stellar and planetary
magnetospheres lead to magnetic stresses that may result in reconnection
and reconfiguration events where energy is eventually dissipated by
high-energy radiation.

As a possible test of this scenario we focused our attention on
\objectname{\hd},
a system with a Jupiter-mass planet in an eccentric orbit. 
In fact, SPI is expected 
to have a strong dependence on the star--planet separation \citep{Cuntz2000}, 
and in systems with elongated orbits
it must occur preferentially or exclusively in proximity of periastron, 
when the stellar and planetary magnetospheres are close 
enough to interact.

\section{Target and Observations}
\label{sec:data}

The system is composed by a G0 primary ($B-V = 0.64$) at 75\,pc from the Sun
and a transiting planet
($M_{\rm p} \sim3.2M_{\rm J}$) on a 21.2 days period orbit with eccentricity 
$e = 0.68$, and periastron and apoastron distances
of 7.4 R$_*$ and 39.1 R$_*$, respectively 
\citep[semimajor axis $a= 0.16 $AU,][]{Barbieri2009}. 
The host star 
is characterized by a chromospheric activity index $\log R'_{\rm HK}$
ranging from $-5.06$ to $-5.01$ \citep{Pace2013},
typical of middle-aged solar-type stars.

\section{Coordinated X-ray and optical campaign}
\label{sec:a+r}

We observed \hd\ with the EPIC camera on board \xmm\ in two visits
of $\sim 30$\,ks each, on September 5 and September 20 2014.
The aim was to observe the star with the planet at the periastron passage
and far from periastron in order to seek phase-related X-ray variability
which could be ascribed to SPI.

The Observation Data Files (ODF) were reduced with SAS ver. 13.5 and the
latest set of calibration files (CCF). For each exposure, we obtained tables of the
events calibrated in detector positions, arrival time, energy, event {\em PATTERN} and
quality flag.
We filtered the events for the recommended flags and patterns
({\sc FLAG=0}, {\sc PATTERN $\le 12$}), with energies in the
broad band 0.3--8.0\,keV, and in a soft band 0.3--1.5\,keV.  
Since a soft spectrum is expected for low-activity stars like \hd,
we preferred the soft band for source detection in the EPIC images,
as already experimented with the observations of HD\,189733,
having the bulk of X-ray emission below 1.5 keV 
\citep{Pillitteri2014}.

For the detection of the star and for estimating
significance and rate, or upper limits, we used a code based on a multi-scale 
wavelet convolution \citep{Damiani1997a,Damiani1997b}.

\hd\ was also observed with the HARPS-N spectrograph \citep{Cosentino2012}
at the Italian 3.6m Telescopio Nazionale Galileo.
In this paper we present data acquired between 2014 August and 
December, in preparation and as a follow-up of the \xmm\ program.
The full observation log is shown in Tab.\,\ref{tab:obslog}. 
Note in particular two HARPS-N observations coordinated with \xmm,
on 2014 September 9 and 21, the latter effectively
simultaneous with \xmm\ while the planet was at periastron. 

Optical spectra were acquired with different exposure times, for different aims:
short exposures (15 minutes each) belong to a long-term series aimed to
precise radial velocity measurements, while long exposures
(120 or 180 minutes) were optimized for high-resolution spectroscopy
of the Ca II H\&K 3968.5, 3933.7\,\AA\ chromospheric emission lines. 
In order to avoid saturation at longer wavelengths, the long exposures were
always split in a number of 15 minutes sub-exposures, and
the raw images were combined a posteriori to compute an average spectrum 
using the standard HARPS-N pipeline. 
In this respect, we have followed the same observation strategy
and data reduction and analysis detailed in \citet{Borsa2015}.

These data were employed to update the orbital parameters. Adopting the
transit timing reported in \citet{Nutzman2011}, we obtained the
following ephemeris ($BJD_{\rm TDB}$) for the passage at periastron:
\begin{equation}
\label{eq:ephem}
T_{\rm 0} = (2454884.3106 \pm 0.0050)
+(21.216398 \pm 1.6\times10^{-5}) E.
\end{equation}
The orbital phases at the start of each observation are reported in
Tab.\ref{tab:obslog}.

\begin{figure*}
\centering
\includegraphics[width=0.80\textwidth]{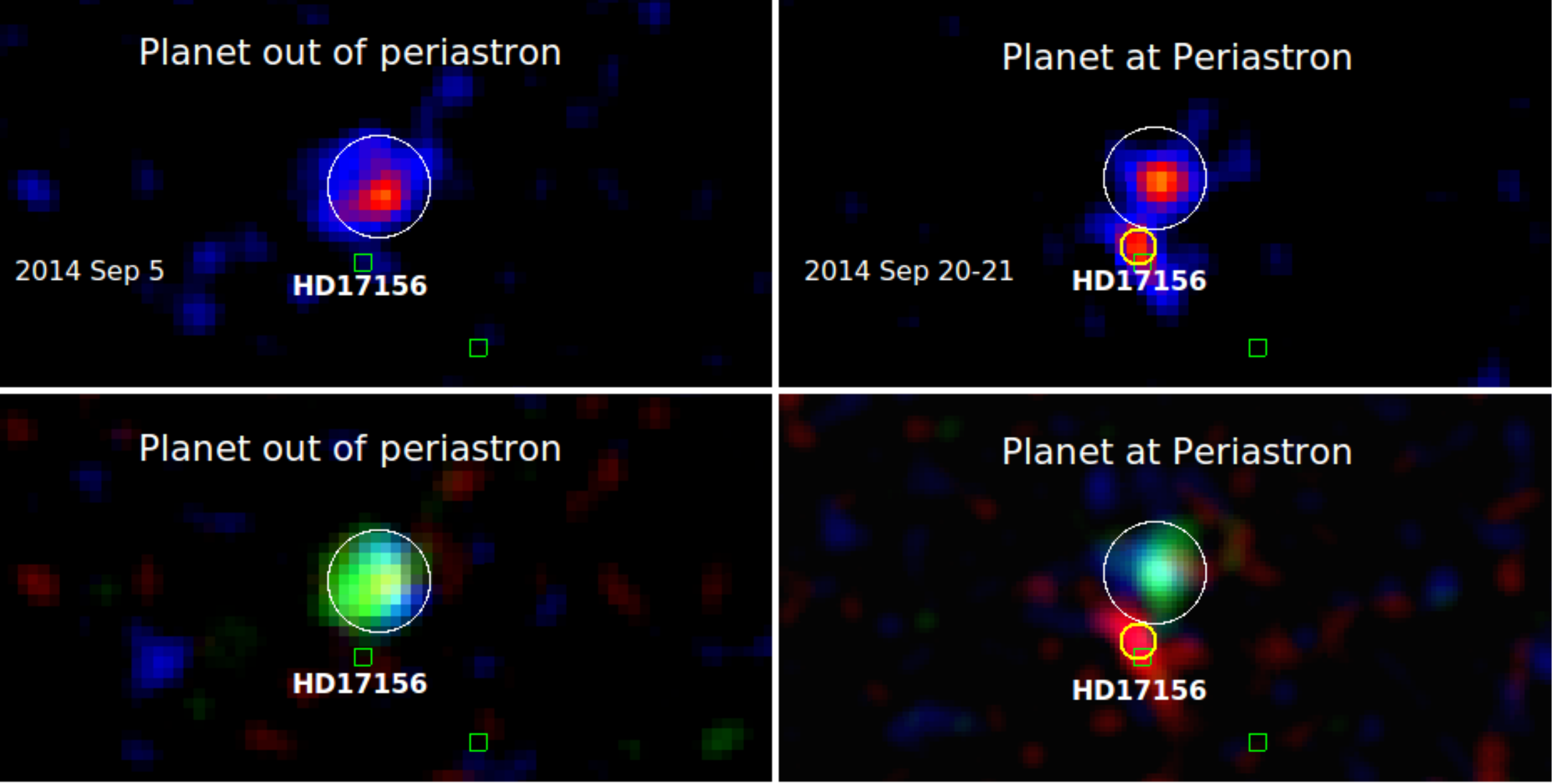}
\caption{\label{ximag}
X-ray images of \hd\ taken far from the planetary periastron (left side
panels) and near the periastron (right side panels).
Top row panels are intensity images, bottom row panels are RGB images with 
photon energy colors $R=0.3-1.0$ keV, $G=1.0-2.5$ keV, $B=2.5-5.0$ keV.
Smoothing is applied to the images, with a Gaussian of $\sigma = 2.4 \arcsec$
for the intensity images and  $\sigma = 4 \arcsec$ for the RGB images.
Positions of the only two objects in the SIMBAD catalog are shown 
with small squares. 
Circle sizes indicate the wavelet detection scales of \hd\ and of an
unrelated background object with an harder spectrum.}
\end{figure*}

\section{Results}
\subsection{X-Ray Emission Switching On}
In Fig. \ref{ximag} we show the field of view around \hd\ in the two \xmm\
exposures. The brightest source visible in both images is an unidentified background
source, at $\sim 10.5"$ from the optical position of our target.
\hd\ was not detected while the planet was approaching the
apoastron (September 5), at a level of $3\sigma$ of the local background,
while it was detected in the visibility window that started 
9 hr after periastron (September 20) and lasted for about 10 hr.
Employing the {\sc pn} and {\sc MOS~2} data\footnote{{\sc MOS~1} data were discarded
because of the presence of a CCD gap near the target location.},
we obtained a $5.5\sigma$ significance level in the soft band, with a {\sc pn}-equivalent
count rate of $(7.1\pm2.5) \times 10^{-4}$\,cts s$^-1$ using a maximum wavelet
spatial scale of $5.6"$. About $26\pm9$  photons were collected
in total in the two EPIC detectors, and their time distribution is
compatible with a constant source (Kolmogorov--Smirnov test probability
$> 50$\%).

The evaluation of the target X-ray flux is hampered by the presence of the
contaminating background source, detected in the first observation with
a count rate $\sim 50$\% higher than in the second observation. 
However, this source is quite hard and
highly absorbed: in fact, its spectrum can be described as a thermal source with
$kT \sim 1.6$\,keV attenuated by a hydrogen column density 
$N_{\rm H} = 1.3 \times 10^{22}$ \nhu.
Hence, we employed only the events collected with the {\sc pn} detector
in the soft band (0.3-1.5 keV) already used for the source detection process. 
Taking into account 
the Encircled Energy Fraction for a point source
within a $5.6"$ radius,
we obtain a count rate of $(4.8\pm2.5) \times 10^{-4}$ cts/s for \hd.
We estimated the source X-ray flux assuming optically thin emission from a
thermal plasma with solar composition and $T = 2$\,MK, as appropriate for
a solar-type corona, thus obtaining $f_{\rm x} = 5.7 \times 10^{-16}$ \fu.
A similar procedure yielded an upper limit $f_{\rm x} < 3.3\times10^{-16} $ \fu for the
source emission far from the periastron.

In order to evaluate the intrinsic source X-ray luminosity, we estimated the
interstellar absorption starting from the stellar effective temperature
$T_{\rm eff} = 6082 \pm 60$\,K \citep{Gilliland2011}, which implies an
unreddened color $(B-V)_0 = 0.563 \pm 0.017$ \citep{sekiguchi+fukugita00}, 
and a color excess $E(B-V) = 0.08$, yielding
an optical extinction $A_{\rm V} = 0.24$, 
and hence $N_{\rm H} \sim 5 \times 10^{20}$ \nhu \citep{guver+ozel2009}.
From this value of absorption and the thermal emission model assumed
above,
we obtain an unabsorbed flux of $f_{\rm x} = 1.5\times10^{-15}$ \fu, and a corresponding X-ray 
luminosity of $L_x \sim 1.0\times10^{27}$ \lu\ close to periastron. 
Similarly, far from periastron the upper limit is $L_x \le
6.0\times10^{26}$ \lu.
We note however that these estimates are very sensitive to the assumed 
plasma temperature. For example, a 1 MK plasma would imply 
X-ray luminosities 6 times larger than reported above.

\begin{figure}
\centering
\includegraphics[width=\hsize]{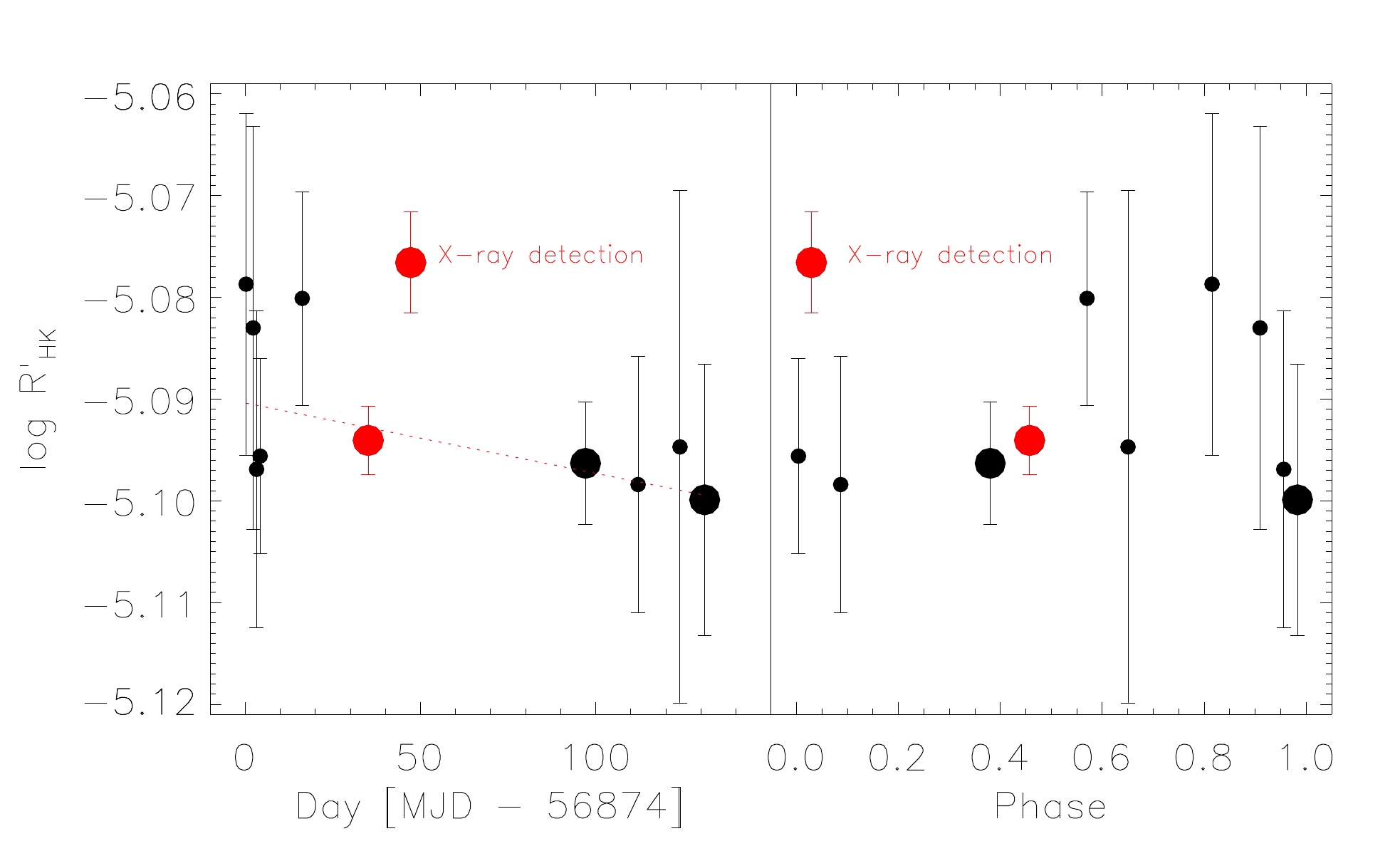}
\caption{\label{fig:caII}
Ca II H\&K chromospheric emission index with $3\sigma$ error bars vs.\
time (left) and vs.\ phase (right).
Symbol sizes indicate different exposure lengths.
In red the HARPS-N observations coordinated with \xmm, and the
baseline obtained with an error-weighted regression, 
excluding only the data point which corresponds to the X-ray detection of the target.
}
\end{figure}

\newpage

\subsection{Chromospheric emission}
The Mount Wilson $R'_{\rm HK}$ index was computed by the
HARPS-N pipeline following the recipe of \citet{Lovis2011}. We
recall that this is a measure of the chromospheric emission in the 
Ca II H\&K lines, corrected for a basal level depending on the stellar
$B-V$ color, and normalized to the stellar bolometric luminosity.
In Fig. \ref{fig:caII} we show the variation of $\log R'_{\rm HK}$ vs.\ time,
from 2014 August to December, and vs.\ orbital phase. 
In the 5 months considered, the data suggest a 
chromospheric activity level essentially flat within the uncertainties
of the measures.
Most remarkable is the significantly higher ($> 8\sigma$)
value of $R'_{\rm HK}$ measured simultaneously with the X-ray detection of
our target. In fact, we observed a 4\% increase of the chromospheric
emission at phase 0.03 (just after periastron) with respect to the previous
measurement taken at phase 0.46 (approaching apoastron).
High $R'_{\rm HK}$ values apparently occurred also on other dates in
2014 August, but these measurements are
affected by relatively large uncertainties because they were
derived from observations with short exposures. Some low-level
variability could also be due to residual systematics in the
reduction of the HARPS-N data \citep{Lovis2011}.

\section{Discussion and conclusions}
\label{sec:discuss}
\hd\ showed enhanced chromospheric and coronal emission a few hours 
after the passage of the planet at the periastron. 
Both the X-ray luminosity and the
$R'_{\rm HK}$ index indicate a star with a
magnetic activity level lower than expected. In fact,
modeling of the Rossiter--McLaughlin effect yields a
rotational velocity of $4.2 \pm 0.3$\,km s$^-1$ \citep{Narita2009}, and
the computed rotation period $P_{\rm rot} \sim 18$\,days leads to a
presumed X-ray luminosity $L_{\rm x} = 1$--$2 \times 10^{28}$\lu,
adopting the activity--rotation relations by \citet{Pizzolato2003}.
Starting from a stellar age of $3.37_{-0.47}^{+0.20}$\,Gyr, based on
transit and asteroseismic observations by \citet{Nutzman2011}, we also
derive $R'_{\rm HK} \sim -4.8$ and $L_{\rm x}/L_{\rm bol}$ in the range
$[-5.8,-6.0]$
from the activity--age relations by \citet{Mamajek2008}. Our observed
chromospheric and coronal emission levels are, respectively, 
a factor $\sim 2$ and $\sim 10$ lower than the values above. 
On the other hand, the measured chromospheric index indicates an X-ray
luminosity 1--2$\times 10^{27}$\lu, in agreement with the \xmm
observations. We conclude that \hd\ was in a low-activity state, with a
``quiescent'' X-ray luminosity just below the detection sensitivity
of the Sept 5 \xmm observation. Consequently, in the following we
assume an average large-scale magnetic field strength of just 1\,G
\citep{Vidotto2014}, and an energy release $\Delta L_{\rm x} \sim 5 \times 10^{26}$\lu
due to an SPI event at the periastron. We deem unlikely the
occurrence of a spontaneous flaring event of such power 
($\Delta L_{\rm x} \approx$ quiescent \lx) in the corona of a quiet star,
such as \hd, just after reaching the minimum star--planet separation. 

One hypothesis is that the SPI phenomenon
has a magnetic origin. We evaluated the energy budget with
the magnetic SPI models by \citet{Lanza2013}.
We assume a loop interconnecting the stellar magnetic field
with a planetary field of $\sim 10$~G
\citep[based on analogy with our Jupiter and theoretical
models, e.g.][]{Reiners2010}. For magnetic confinement, the plasma
temperature near the planet position should not exceed
0.7~MK. In this case, the maximum available power
due to magnetic stresses near periastron is 
$P_{\rm stress} \sim 2 \times 10^{27}$ erg s$^{-1}$, 
for a stellar potential dipolar configuration.
We do not expect any energy release by this mechanism at apoastron. 
A possible difficulty with this explanation rests on the MHD time scales
for establishing an interconnecting loop, building up the magnetic
stresses and releasing the accumulated energy, given the fast planetary
passage. In fact, the relative velocity of the magnetic field
co-rotating with the star with respect to the planet speed at periastron
is about 160\,km s$^{-1}$, implying a dynamical time scale of 10--15
hr. Stability of such a long loop ($l \sim 10^{12}$\,cm) is also an
open issue.
Reconnection between separate stellar and planetary
magnetospheres is also unlikely, because the dissipated power would be lower 
than estimated above by about four orders of magnitude \citep{Lanza2012}.

An alternative possibility is
that the power emitted in X-rays is released by matter evaporated
from the planet and accreted onto the star near periastron
\citep{Matsakos2015, Pillitteri2015}. 
This requires a mass accretion rate larger than 
$\sim 3.6 \times 10^{11}$\,g s$^{-1}$ to make a power of 
$5 \times 10^{26}$\,erg s$^{-1}$ available.
The free-fall time from the inner Lagrangian point L1 to the star is
about 14 hr, making this process compatible with the delay of the
brightening we observed after periastron. On the other hand, detailed
simulations are required to understand the mechanism and time scale of
accumulation and spill over of evaporated material onto the star.

In any case, inspection of Fig.\ \ref{fig:caII} shows that
no significant chromospheric brightening was observed near 
other periastron passages, preceding or following the one
highlighted by the X-ray detection. We conclude that SPI events in this
high-eccentricity system occur occasionally, 
or they have a short time duration.
Monitoring of \hd\ will be continued in the frame of the GAPS programme
with the aim to better understand the frequency of the putative SPI
effects and their nature. On the other hand, new X-ray observations
are severely limited by the faintness of the target and by visibility
constraints with \xmm.

\acknowledgements{AM and the GAPS project team acknowledge support from
the ``Progetti Premiali'' funding scheme of the Italian Ministry of
Education, University, and Research.
IP acknowledges support from the European Union Seventh
Framework Programme FP7/2007--2013, under the grant agreement n$_{\rm
o}$ 267251 ``Astronomy Fellowships in Italy'' (AstroFIt).
}

{\it Facilities:} \facility{\xmm (EPIC)}, \facility{TNG (HARPS-N)}

\bibliographystyle{apj} 

\end{document}